\documentstyle[12pt,epsfig]{article}
\author{Esteban Roulet$^a$, Guenter Sigl$^b$, Arjen van Vliet$^b$, Silvia Mollerach$^a$ \\
$^a$ {\it CONICET, Centro At\'omico Bariloche,}\\
{\it Av. Bustillo 9500 (8400) Argentina.}\\
$^b$ {\it II. Institut f\"ur Theoretische Physik, Universit\"at Hamburg,}\\
{\it Luruper Chaussee 149, 22761 Hamburg, Germany}}

\title{PeV neutrinos from the propagation \\ of ultra-high energy cosmic rays}

\begin{document}

\maketitle

\begin{abstract}
 We discuss the possibility that the PeV neutrinos recently observed by IceCube 
are produced by the interactions of extragalactic cosmic rays during their propagation through the radiation backgrounds.
We show that the fluxes resulting from the decays of neutrons produced in the interactions of cosmic ray protons with the CMB background are suppressed ($E_\nu^2$d$\Phi_\nu/$d$E< 10^{-10}$ GeV/cm$^2$ s sr), with those resulting from the decays of pions produced in the interactions with the UV/optical/IR backgrounds being the dominant ones at PeV energies.
The anti-neutrino fluxes produced by the decay of neutrons resulting from the photodisintegration of heavy nuclei with CMB photons are also shown to be quite suppressed ($E_\nu^2$d$\Phi_\nu/$d$E< 10^{-11}$ GeV/cm$^2$ s sr), while those produced by photo-pion processes with UV/optical/IR backgrounds may be larger, although they are not expected to be above those achievable in the pure proton case. Scenarios with mixed composition and low cutoff rigidities can lead to PeV neutrino fluxes enhanced with respect to those in the pure Fe scenarios. We also discuss the possible impact of the Glashow resonance for the detection of these scenarios, showing that it plays a moderate role.

\end{abstract}

\section{Introduction}
The IceCube Collaboration recently reported \cite{is12} the observation of two neutrinos with energies in the range 1--10~PeV, which are the highest energy neutrino events observed up to now. Possible sources for their origin include atmospheric neutrinos (in particular from the prompt charm production), astrophysical neutrinos produced by photo-pion or $pp$ interactions of accelerated protons with ambient radiation or matter in sources such as gamma ray bursts, active galactic nuclei or starburst galaxies, or cosmogenic neutrinos produced during the propagation of ultra-high energy cosmic rays (UHECR) through the extragalactic radiation backgrounds. 

Although two events are certainly not enough to determine the neutrino flux, these observations suggest a flux level of $E_\nu^2$d$\Phi_\nu/$d$E \simeq 10^{-8}$~GeV/cm$^2$s~sr in the 1--10 PeV range \cite{is12}. Note that the canonical expectations for the rapidly falling atmospheric background neutrino flux  are about one order of magnitude below this level, while astrophysical source scenarios may produce the required flux. It is our purpose here to discuss  some generic upper bounds on the cosmogenic neutrino fluxes of various origins to identify the potentially dominant contributions at PeV energies.  

Another relevant feature of the observations is that the two neutrinos are in the cascade mode (i.e. not involving muon or tau tracks), so that they are due to either a NC interaction of any of the three neutrino (or antineutrino) flavors 
or to charged current interactions of electron neutrinos (or antineutrinos). One interesting aspect of this is that the $\bar\nu_e$ channel presents, besides the interaction with nucleons, a resonant interaction with the electrons at an energy of $E_{GR}=M_W^2/2 m_e=6.3$~PeV (the Glashow resonance \cite{gl60}), and we will also discuss the possible impact of this channel for the IceCube detection.

\section{Neutrino fluxes from UHECR protons}

Extragalactic UHECR protons with energies above $6\times 10^{19}$~eV get attenuated when propagating through the cosmic microwave background (CMB) mostly by pion production processes, which lead to the well known expected GZK suppression \cite{gzk}. These losses reach a maximum strength  at the $\Delta(1232)$ resonance, i.e. when $m_\Delta^2\simeq m_p^2+ 2 E_pE_\gamma(1-\cos\theta)$, with $\theta$ the angle between the $p$ and $\gamma$ momenta in the lab frame. This corresponds to a proton energy (adopting $\theta=\pi$) of $E_p(\Delta)\simeq 1.6\times 10^{20}$~eV$/(E_\gamma/10^{-3}$~eV), with  the average CMB photon energy being $\sim 0.7\times 10^{-3}$~eV at present (i.e. for redshift $z=0$), see \cite{be90} for a comprehensive review.  The process $p\gamma\to \pi^{+}n$ leads to the production of cosmogenic neutrinos \cite{be69} both through the pion decay chain $\pi^+ \to \mu^+\nu_\mu\to e^+ \nu_e\bar\nu_\mu\nu_\mu$ and through the neutron decay $n\to pe\bar\nu_e$ \cite{en01}. Since the pions typically carry about 1/5 of the proton energy, each neutrino from the pion decay has on average an energy of about $E_\nu\simeq E_p/20$.    On the other hand, the $\bar\nu_e$ from the neutron decay has a typical energy of $E_\nu\simeq 4\times 10^{-4} E_n\simeq 3\times 10^{-4} E_p$.

When these interactions happen at high redshift, since the CMB temperature scales as $(1+z)$  the proton energies for which the photopion production start to be efficient are $E_p\simeq E_{GZK}/(1+z)$, where for definiteness we adopt $E_{GZK}=10^{20}$~eV as the typical proton energy for pion production at $z=0$. Note also that the CMB photon density increases as $(1+z)^3$, making the opacity of the universe to UHE protons correspondingly higher.
The energies of the neutrinos produced at high redshifts get further reduced by the adiabatic losses as they propagate to us, leading to $E_\nu\simeq E_{GZK}/(20(1+z)^2)$ for the neutrinos originating from pion decays, and $E_\nu\simeq 3\times 10^{-4} E_{GZK}/(1+z)^2$ for those from neutron decays.  The neutrino production turns out to be sizeable up to redshifts of 3--5 \cite{en01}, depending on the actual source redshift evolution, and for instance considering a typical neutrino production redshift $z\simeq 1.2$ one gets peaks on the neutrino spectrum resulting from interactions with the CMB at energies $\sim 10^{18}$~eV (from $\pi$ decays) and $\sim 6\times 10^{15}$~eV (from $n$ decays). These peaks are anyway quite wide, because the $\Delta$ resonance is wide (and other pion production channels contribute as well), because the CMB photons have a wide thermal spectrum and also because different redshifts contribute to the neutrino production.

A useful relation can be obtained between the two neutrino fluxes just considered, since in the charged pion producing interactions the same number of low energy $\bar\nu_e$ and higher energy $\nu_e,\ \bar\nu_\mu$ or $\nu_\mu$ are produced. Hence, denoting by $\Phi_\nu$ the resulting neutrino diffuse fluxes, one has,  for the fluxes produced in interactions with the CMB alone, that
\begin{equation}
\left[\frac{{\rm d}\Phi_{\bar\nu_e}}{{\rm d}{\rm log}E}\right]_{(E_{\bar\nu_e}=6\times 10^{15}\ {\rm eV})}^{n-dec, CMB} \simeq   \left[ \frac{{\rm d}\Phi_{\nu_\mu}}{{\rm d}{\rm log}E}\right]_{(E_{\nu_\mu}= 10^{18} \ {\rm eV})}^{\pi-dec, CMB}.
\end{equation}

Using that the EeV neutrinos are actually dominated by those produced in interactions with the CMB (see below), and ignoring for the time being the effects of the neutrino oscillations on the $\bar\nu_e$, we then get 
\begin{eqnarray}
\left[E_\nu^2\frac{{\rm d}\Phi_{\bar\nu_e}}{{\rm d}E}\right]_{(E_{\bar\nu_e}=6\times 10^{15}\ {\rm eV})}^{n-dec, CMB}
& \simeq &  6\times 10^{-3} \left[E_\nu^2 \frac{{\rm d}\Phi_{\nu_\mu}}{{\rm d}E}\right]_{(E_{\nu_\mu}= 10^{18} \ {\rm eV})}^{\pi-dec, CMB} \cr
& \simeq &  2\times 10^{-3} \left[E_\nu^2 \frac{{\rm d}\Phi_{{\rm all}\ \nu}}{{\rm d}E}\right]_{(E_{\nu}= 10^{18} \ {\rm eV})} .
\label{bound1}
\end{eqnarray}

The  all flavor diffuse neutrino flux has been constrained at EeV energies by the unsuccessful searches by IceCube  \cite{icnubound} and Auger \cite{augernubound}, which imply the approximate bound $E_\nu^2 $d$\Phi_\nu/$d$E < 3\times 10^{-7} $GeV/cm$^{2}$ s~sr at EeV energies. 
Moreover, in proton scenarios a stronger bound has been obtained indirectly from the so-called cascade decays \cite{be75}.
  This bound is derived from the requirement that  the $\pi^0$ decay gammas (also produced in the photo-pion processes) and the $e^+ e^-$ pairs do not produce too large amounts of GeV--TeV photons when they cascade down to low energies as they interact with the CMB and IR radiation backgrounds. The allowed amount of low energy photons is bounded by  the diffuse photon background measured by the Fermi LAT experiment \cite{ab10}. The cascade limit which results  for the all flavor cosmogenic neutrino flux at EeV energies
is at the level of  $E_\nu^2 $d$\Phi_\nu/$d$E < 5\times 10^{-8} $GeV/cm$^{2}$ s~sr \cite{be11,ah10}. 
This, combined with eq.~(\ref{bound1}), implies that the $\bar\nu_e$ flux at PeV energies produced from interactions with CMB photons should satisfy  $E_\nu^2 $d$\Phi_{\bar\nu_e}/$d$E < 10^{-10} $GeV/cm$^{2}$ s~sr. This upper bound is about two orders of magnitude below the flux level suggested by the two PeV neutrinos observed by IceCube, and hence can hardly be responsible for those events (this is at variance with the interpretation suggested in \cite{ba12}).

Cosmic ray protons with energies below $6\times 10^{19}$~eV (at $z=0$) loose energy mainly by $e^+e^-$ pair creation interactions with CMB photons, without producing neutrinos. However, below $\sim 10^{18}$~eV the dominant attenuation process becomes the photopion production in interactions with UV, optical and IR radiation backgrounds, which being more energetic than CMB photons lead to a reduced proton energy to produce the $\Delta$ resonance ($E_p(\Delta) \simeq 1.6\times 10^{17}$~eV$/(E_\gamma/{\rm eV})$). In this case, the neutrinos produced by the pion decays  have energies $\sim 8\times 10^{15}$~eV/[$(1+z)(E_\gamma/{\rm eV})]$, and hence will contribute in the PeV range, while those from the corresponding neutron decays will be at much lower energies ($< 10^{14}$~eV), and hence buried below the atmospheric neutrino background. The PeV neutrino fluxes from UHECR proton scenarios are indeed dominated by those from pion decays produced in interactions with UV/optical/IR photons. Numerical simulations show that  all flavor neutrino flux levels of $E_\nu^2 $d$\Phi_\nu/$d$E  \simeq few\times 10^{-9} $GeV/cm$^{2}$~s~sr
can be achieved  for $E_\nu\simeq{\rm PeV}$ (see e.g. \cite{st05,al06,ta09,ko10}), with the precise values depending on the assumed source evolution, on the radiation background adopted  and on the shape of the extragalactic proton spectrum at the source. 

\begin{figure}[t]
\centerline{\epsfig{width=3.5in,angle=-90,file=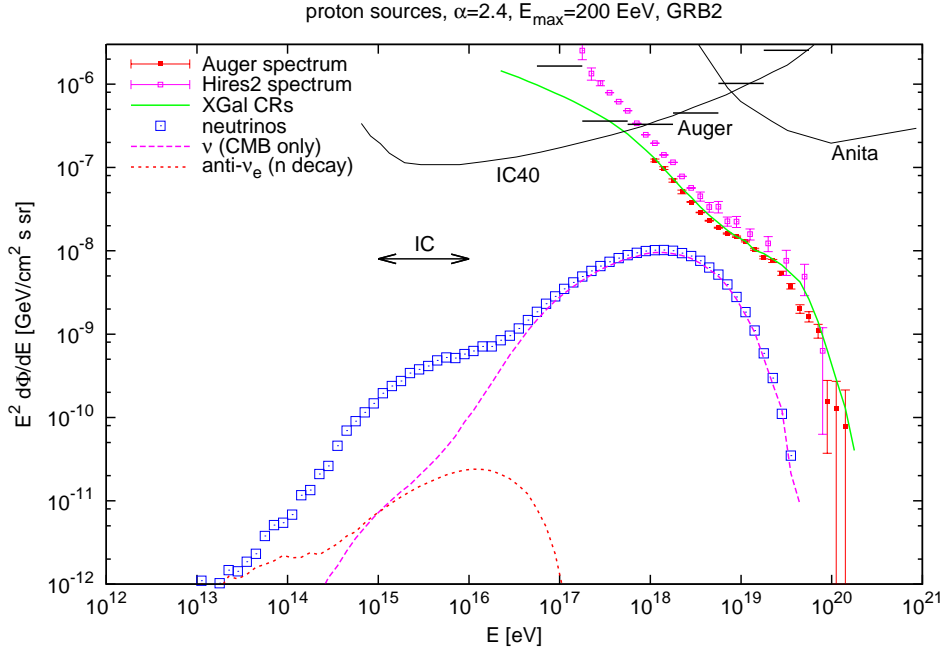}}
\vskip 1.0 truecm
\caption{Proton `dip' scenario with source spectral index $\alpha=2.4$ and $E_{max}=200$~EeV. Indicated are the propagated proton spectrum and the resulting all flavor neutrino fluxes (obtained with CRPropa). We also show separately the neutrino backgrounds due to interactions with CMB alone as well as those resulting from $n$ decays. The CR flux measured by Auger and Hires  and the neutrino limits from IceCube, Auger  and Anita  are displayed. We also indicate the energy range and approximate flux level suggested by the two observed IceCube events.} 
\label{fpgrb2}
\end{figure}

To illustrate this, we show in  figure~\ref{fpgrb2} the  neutrino fluxes resulting in pure proton scenarios, adopting a source power spectrum $\alpha=2.4$, maximum energy of 200~EeV and minimum energy of $2\times 10^{16}$~eV. We also adopt a source redshift evolution (for the density times CR emissivity) following the gamma ray bursts one (corresponding to the SFR6 model derived in ref.~\cite{le07}, here referred to as GRB2). For the  UV/optical/IR radiation background we consider (in all the figures) the one  following the `best fit model' in ref.~\cite{kn04} (including also its redshift evolution).
We obtain the CR spectra as well as the diffuse neutrino fluxes 
using the simulation code CRPropa \cite{crpropa}. Besides the total fluxes, we show separately the contributions resulting from the interactions with CMB photons and that of antineutrinos from $n$-decays alone. The relation obtained in eq.~(\ref{bound1}) can be seen to hold by comparing the heights of the two peaks (the EeV all flavor one and the PeV one from $n$-decays). 

The resulting CR spectrum is normalized at 10~EeV to that measured by the Auger Collaboration \cite{augerspect} (which has a 22\% systematic uncertainty on the energy scale,  not shown in the plot). The overall shape of the spectrum is found to be in reasonable agreement with the measured one above the ankle region. We also display the spectrum measured by the Hires Collaboration \cite{hiresspect}, and  the bounds on the all flavor neutrino fluxes obtained by IceCube \cite{icnubound}, Auger \cite{augernubound} and Anita \cite{anitanubound}.

The PeV neutrino flavors produced in these scenarios, arising mostly from pion decays,  are in the ratio  $(\nu_e:\nu_\mu:\nu_\tau)=(1:1:0)$
 and  $(\bar\nu_e:\bar\nu_\mu:\bar\nu_\tau)=(0:1:0)$. These ratios will  then be affected by the incoherent flavor oscillations that take place from their production until their detection on Earth. For instance, if one adopts the tri-bi maximal (TBM) neutrino mixing pattern, these fluxes get transformed into (0.78:0.61:0.61) and (0.22:0.39:0.39) upon arrival on Earth \cite{pa08}.  Departures from the TBM mixing predictions, as required by the recent measurement of a non-vanishing $\theta_{13}$ \cite{an12}, will induce small shifts on the above mentioned flavor ratios. Note that in the case of $\bar\nu_e$ production in neutron decays,  the initial flavor ratios of the antineutrinos (1:0:0) would change into (0.56:0.22:0.22) by oscillations in the TBM scheme.

In this 'dip' scenario the ankle results from the pair production interactions off the CMB, and for this to work the spectral slope at the sources should be close to $\alpha=2.4$ for the source evolution considered here, and could be even harder ($\alpha\simeq 2.2$) for a stronger source evolution such as that following the evolution of the Faranoff Riley AGNs. We note that for a given source evolution, steepening the source spectra could lead to an enhanced PeV neutrino flux, but would affect the resulting CR flux which will no longer match the observed one at the highest energies.  Regarding the higher energy cutoff, increasing it would have the effect of slightly modifying the details of the CR GZK suppression and slightly widen the EeV neutrino peak (extending it to higher energies), but will have little impact on the PeV neutrino fluxes.

\begin{figure}[t]
\centerline{\epsfig{width=3.5in,angle=-90,file=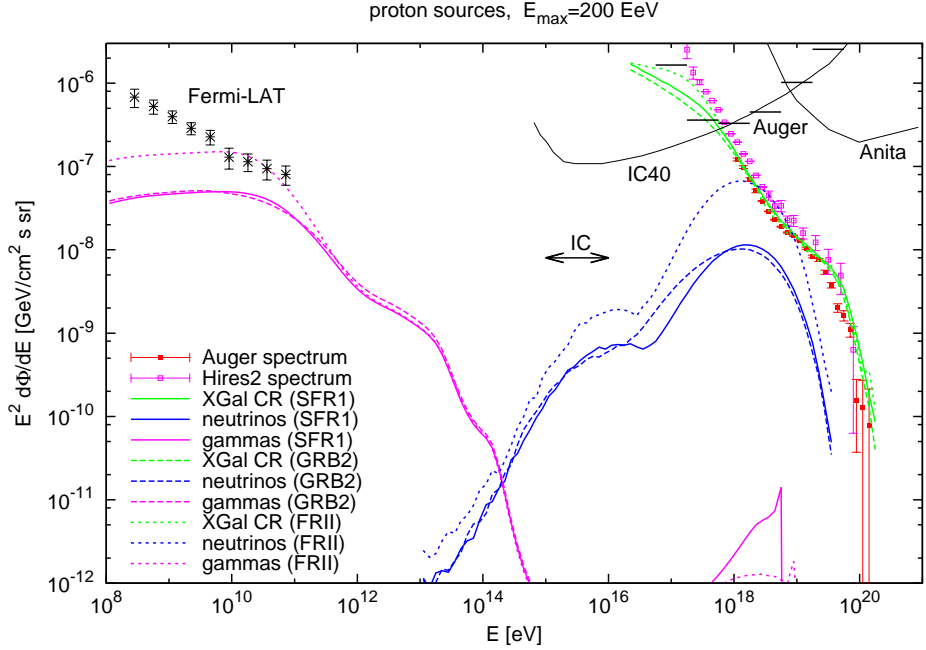}}
\vskip 1.0 truecm
\caption{Proton  scenario with $E_{max}=200$~EeV for different source evolution models (SFR1, GRB2 and FRII). The  source spectral index is $\alpha=2.4$ for the SFR1 and GRB2 models, while $\alpha=2.2$ for the FRII model. Indicated are the propagated proton spectrum, the resulting (all flavor) neutrino and the photon fluxes. The photon background measured by Fermi-LAT \cite{ab10} is indicated, besides the CR spectra and $\nu$ bounds included in fig.~1. } 
\label{fpgam}
\end{figure}

In figure \ref{fpgam} we show the predicted CR spectra, the all flavor neutrino fluxes as well as the cascade photons for three different source evolution  models. These models  follow  the star formation rate (SFR1 in \cite{le07}), the GRB2 model considered in fig.~1 and the AGN model evolution from ref.~\cite{wa05} (FRII). In order to obtain a reasonable fit to the high energy cosmic ray spectrum, we adopt $\alpha=2.4$ for the SFR1 and GRB2 models, while $\alpha=2.2$ for the FRII one. A significant dependence of the predictions on these parameters is observed. We see  that the gamma ray fluxes obtained in the scenario with 
the strongest neutrino fluxes are 
on the verge of conflicting 
with the diffuse gamma ray flux observed by Fermi \cite{ab10}. 

One further issue is that the height of the PeV neutrino peak (resulting mostly from interactions of  $\sim 10^{17}$~eV protons with UV/optical/IR photons) and that of the EeV peak (resulting from interactions of $\sim 10^{20}$~eV protons with CMB photons) are in principle related, depending on the underlying proton spectral shape.  
Hence, the bounds on the EeV diffuse neutrino fluxes (direct or from cascade decays)  may also constrain the allowed maximum height of the PeV neutrino peak. 
  We note however that if the proton spectrum above a few EeV (or even a few tens of EeV) were  suppressed, for instance due to the existence of a low maximum rigidity in the  acceleration process at the sources, the EeV neutrino peak would be suppressed but not the PeV peak, and hence in these cases no constraint from the associated cascade decays would result. 

Of interest in this respect is the so-called disappointing model \cite{al11}, which actually has an enhanced proton flux at EeV energies, with a cutoff at few EeV, and in this scenario the cosmic rays above the ankle at $\sim 4$~EeV are nuclei with increasingly heavier masses. This model would predict an enhanced neutrino flux at PeV energies with no sizeable neutrino peak at EeV energies (see next Section). 

We also note that the PeV neutrino flux is sensitive to the amount of UV photons, and hence scenarios with an enhanced UV background, as considered in \cite{al06,ko10}, can lead to some enhancement in PeV neutrino fluxes. 

\section{Neutrino fluxes from UHECR nuclei}

Scenarios in which heavier nuclei make a significant contribution to the UHECR flux are qualitatively different (see e.g. \cite{av05,ho05}). Here the photopion production off the CMB photons only occurs for energies above $\sim A E_{GZK}$, where $A$ is the mass number of the nucleus. On the other hand, photodisintegration processes play an important role at lower energies. Photodisintegrations are dominated by the giant dipole resonance (GDR) which, in the nucleus rest frame, has a threshold for photon energies between a few MeV and 10 MeV (depending on the nucleus), and peaking at about 20 MeV. The photon energy in the CR rest frame can be expressed as
\begin{equation}
E'_\gamma=\frac{E}{Am_p}(1-\beta \cos\theta)E_\gamma \simeq 3.8 \ {\rm MeV}\left(\frac{56}{A}\right)\left(\frac{E}{10^{20}\ {\rm eV}}\right)\left(\frac{1-\beta\cos\theta}{2}\right)\frac{E_\gamma}{10^{-3}\ {\rm eV}},
\end{equation}
where $\beta\equiv\sqrt{1-(m_pA/E)^2}$ is the boost factor and $\theta$ is the angle between the CR and photon momenta in the lab frame. While they photodisintegrate, nuclei emit nucleons without  changing their Lorentz factor, hence just loosing energy because the mass of the leading fragment gets reduced. In this process secondary nucleons  are emitted with energies $E/A$ for energies $E>E_{GDR}\simeq 10^{20}$~eV$A/56$. If the photodisintegrations take place instead at non-zero redshift, the secondary nucleons final energy will be $E/A(1+z)$, where the contributions arise from $E>E_{GDR}/(1+z)$.  About half of these nucleons will be emitted as neutrons, which then decay to produce $\bar\nu_e$ with energies $E_\nu\geq (4\times 10^{-4})(10^{20}{\rm eV}/56)/(1+z)^2\simeq {\rm few}\ 10^{14}$~eV (assuming interactions with the CMB background alone). Note that for interactions with the CMB, only those neutrinos produced  by energetic nuclei at or above the GDR will end up with energies above one PeV.  The nuclei of smaller energies  may also photodisintegrate  by interacting with the UV/optical/IR radiation backgrounds, leading to a flux of  $\bar\nu_e$ of low energy ($< 10^{14}$~eV), and hence are not relevant  here.

One may obtain a conservative upper bound on the flux of the $\bar\nu_e$ produced at PeV energies by noting that the secondary nucleons produced by photodisintegration 
 with CMB photons  end up, regardless of the primary nucleus mass, piled up around a few EeV, and only about half of these were produced as neutrons. Hence, by requiring that the secondary nucleons flux at the relevant energy be below the actually measured CR flux one ends up with
\begin{equation}
\left[E_\nu^2\frac{{\rm d}\Phi_{\bar\nu_e}}{{\rm d}E}\right]_{(E_{\bar\nu_e}\simeq 10^{15}\ {\rm eV})}^{n-dec}
\simeq  \frac{ 4\times 10^{-4}}{2} \left[E^2 \frac{{\rm d}\Phi_{CR}}{{\rm d}E}\right]_{(E= 2.5\times 10^{18} \ {\rm eV})}< 10^{-11} {\rm \frac{GeV}{cm^2\ s\ sr}},
\label{bound2}
\end{equation}
where we used that the measured CR flux at 2.5~EeV is about d$\Phi_{CR}$/d$E\simeq 7\times 10^{-18}$/(EeV cm$^2$ s sr) \cite{augerspect}. These bounds may be slightly modified if one considers that the secondary nucleon energies could be degraded by pair production processes, but at few EeV energies this process does  not have a large impact. Hence, the $\bar\nu_e$ flux produced by decays of neutrons resulting from the photodisintegration of UHECR nuclei are quite suppressed at PeV energies. 

On the other hand,  neutrinos with few PeV energies may be produced by the interactions of extragalactic nuclei with UV/optical/IR photons by photopion production. This would require that the energy per nucleon be about $\sim 20 E_\nu$, and hence the nuclei should have energies of about $20AE_\nu$.  However, since the UHECR source spectrum is expected to be steeper than $E^{-2}$, the photopion contribution from scenarios dominated by nuclei  will not be larger than that arising in the proton dominated scenarios discussed in the previous section (for instance, if all CR had a given mass number $A$, the number of nucleons per logarithmic energy interval at energy $E$ is related to the corresponding number in a pure proton scenario with the same spectral slope  $\alpha$ by the factor $A^{2-\alpha}$). Moreover, the direct production of pions from the nucleons bound inside the nuclei is suppressed with respect to that from free nucleons (see discussion in \cite{al06}). Hence, nuclei scenarios give rise to PeV neutrinos by photopion production of UV/optical/IR photons but at a level which is  not expected to be larger than that achievable in proton scenarios.

\begin{figure}[t]
\centerline{\epsfig{width=3.5in,angle=-90,file=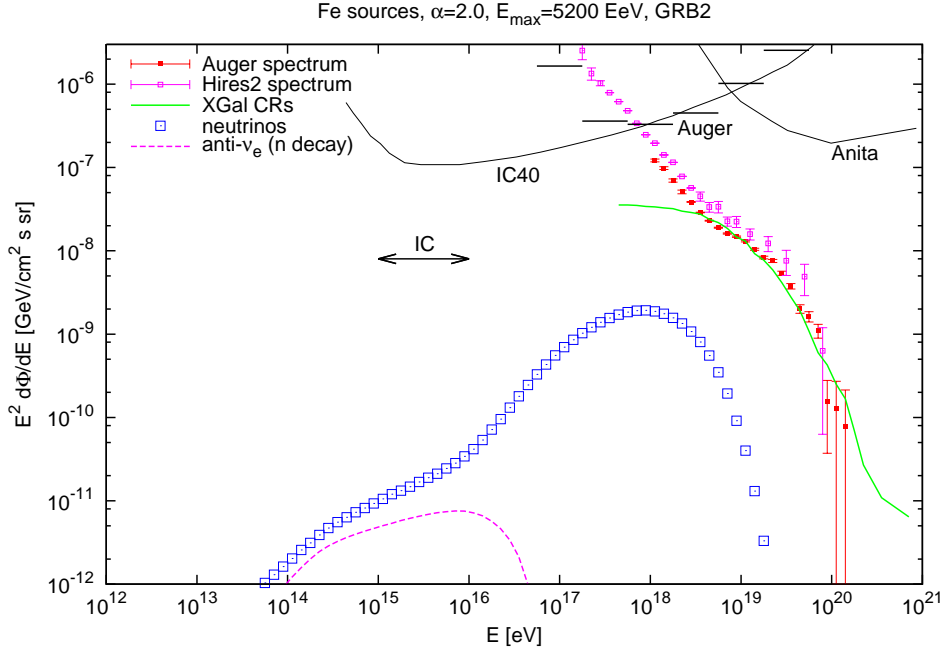}}
\vskip 1.0 truecm
\caption{Extragalactic Fe scenario with source spectral index $\alpha=2.0$ and $E_{max}=5200$~EeV. Indicated are the propagated CR spectrum and the resulting (all flavor) neutrino  fluxes, as well as the neutrino background due to $n$-decays alone. } 
\label{fig2}
\end{figure}

To illustrate the predictions from a heavy composition scenario we show in fig.~\ref{fig2} a case corresponding to Fe only sources with a source spectrum $\alpha =2.0$ and a maximum Fe energy of $5200$~EeV (corresponding to rigidities $R\equiv E/Z<200$~EV, with $Z$ the charge of the nucleus), following the GRB2 source redshift evolution. We also show  separately the neutrino fluxes arising from   the neutron decays, where the bound from eq.~(\ref{bound2}) can be seen to hold. The main contribution to the PeV neutrino fluxes arises from the interactions with UV/optical/IR radiation backgrounds.

We note that the EeV neutrino peak strongly depends on the assumed maximum Fe energy at the source, and considering lower maximum energies can drastically reduce this peak (which essentially disappears for $E_{max}< 1000$~EeV, corresponding to $E/A<20$~EeV). This would however not affect in a significant way the expectations for the PeV neutrinos in these Fe scenarios.

\begin{figure}[t]
\centerline{\epsfig{width=3.5in,angle=-90,file=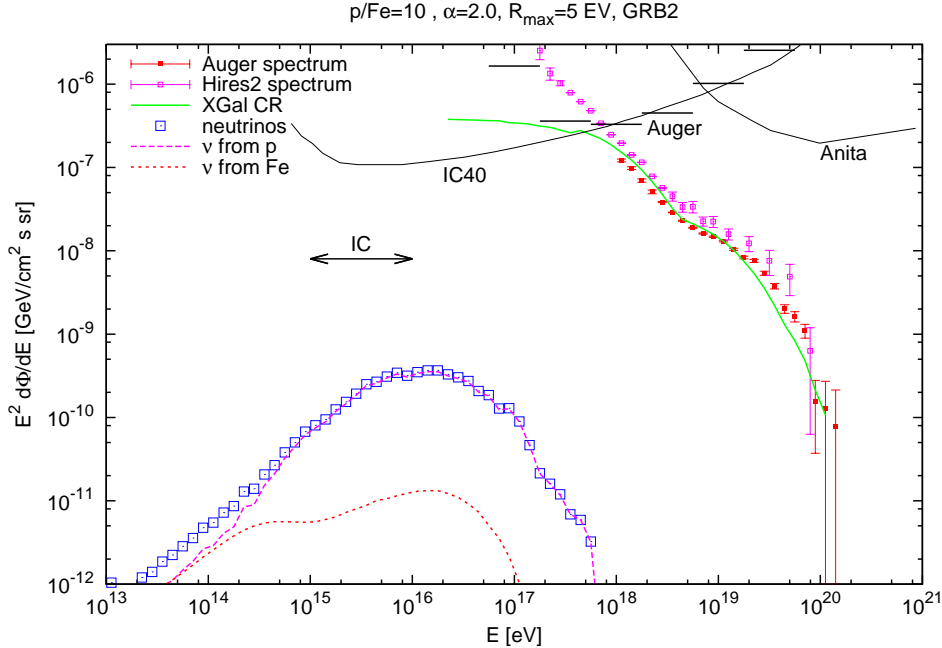}}
\vskip 1.0 truecm
\caption{Mixed composition (p--Fe) scenario with source spectral index $\alpha=2.0$ and $E_{max}=5\,Z$~EeV. Indicated are the propagated CR spectrum, the resulting (all flavor) neutrino  fluxes and the separate contributions from $p$ and Fe primaries. } 
\label{fig3}
\end{figure}

Figure~\ref{fig3} shows instead the results  obtained in a scenario having a mixture of p and Fe and a low energy cutoff ($E_{max}=5 Z$~EeV), inspired in the `disappointing'  model \cite{al11}. The relative source abundances considered are $n_p/n_{Fe}=10$ at a given energy (below the proton cutoff), with power spectrum $\alpha=2.0$ and for the GRB2 source evolution model. We see that the enhanced proton contribution below the ankle helps to reach a larger flux of PeV neutrinos than in the pure Fe case. In this mixed composition scenario the CR spectrum in the ankle region is similar to the measured one, but it does not fit well the highest energies. This may however depend on the precise distribution of nearby sources and on the shape of the source cutoff adopted.

We note that in scenarios with more than two components, e.g. those in which the average CR mass  gradually increases above the ankle, harder spectra for each source component are required to fit the observed overall spectrum, and hence this will tend to reduce the fluxes of PeV neutrinos with respect to those found for the p-Fe only mixture.

\section{The impact of the Glashow resonance}

An interesting aspect of PeV neutrinos is that the $\bar\nu_e$ may be detected through the interactions with electrons \cite{gl60,bh11,xi11}, which gets resonantly enhanced at the $W$ pole, corresponding to an energy $E_{GR}=M_W^2/2m_e=6.3$~PeV. The $W$ decays with $\sim 10$\% branching into each lepton flavor and the remaining 70\% into hadronic channels. The electron and hadronic channels (i.e. about 80\% of the decays) would give rise to a cascade signature, while the muon and tau channels give rise to tracks (or lollipops, since the tau has a track length of about 300~m at 6 PeV, decaying then to produce a cascade). The resonant cross section is (considering e.g. the muonic channel)
\begin{equation}
\sigma^{GR}(\bar\nu_e e \to \bar\nu_\mu \mu)=\frac{G_F^2M_W^2}{3\pi}\frac{E_\nu/E_{GR}}{(1-E_\nu/E_{GR})^2+(\Gamma_W/M_W)^2}.
\end{equation}
The total, i.e. into all fermion final states, cross section reaches a peak value of $5\times 10^{-31}$~cm$^2$, having an associated width  $\Delta=(\Gamma_W/M_W)E_{GR}\simeq 0.17$~PeV, which is quite narrow.

On the other hand, PeV neutrinos of all flavors interact with nucleons through charged current interactions with a cross section \cite{ga98}
\begin{equation}
\sigma^{CC}(\nu_iN\to \ell_iX)\simeq 1.6(E/E_{GR})^{0.363}10^{-33}\ {\rm cm}^2,
\end{equation}
and through neutral currents with 
 \begin{equation}
\sigma^{NC}(\nu_iN\to \nu_iX)\simeq 6.8(E/E_{GR})^{0.363}10^{-34}\ {\rm cm}^2,
\end{equation}
with the cross sections for the antineutrinos having similar values.

To compare the relative importance of the resonant cross section on the rates one has to take into account the relative abundances of the different neutrino and antineutrino flavors, the fact that  the ratio of the electron and nucleon densities in water is $n_e/n_N=5/9$ and that the expected fluxes span a wide energy range. Since, as shown in the previous sections, the dominant fluxes of PeV neutrinos are those resulting from pion decays (produced in interactions with UV/optical/IR photons), the flavor ratios on Earth are  ${\bf \xi}\simeq(0.78:0.61:0.61)$ for neutrinos and   ${\bf \bar\xi}\simeq(0.22:0.39:0.39)$ for antineutrinos (adopting TBM mixing for simplicity). Focusing on the cascade producing channels only, the relative contribution of the GR  channel is, assuming for definiteness that in the energy interval $[E_{min},E_{max}]$ around $E_{GR}$ the flux spectral shape can be approximated as $\propto E^{-2}$,
\begin{equation}
R(E_{min},E_{max})=\frac{n_e}{n_N}\frac{\int_{E_{min}}^{E_{max}}{\rm d}E\ E^{-2}\bar\xi_e 8\sigma^{GR}(\bar\nu_ee\to\bar\nu_\mu\mu)}{
\int_{E_{min}}^{E_{max}}{\rm d}E\ E^{-2}\left(\sum_i(\xi_i+\bar\xi_i) \sigma^{NC}+(\xi_e+\bar\xi_e)\sigma^{CC}\right)}.
\end{equation}
From this expression one gets for instance that for a very narrow interval around $E_{GR}$  with width $\Delta=0.17$~PeV the ratio is
$R(E_{GR}-\Delta,E_{GR}+\Delta)=10.7$, while for an interval with 2.5~PeV width one gets $R(3.8$~PeV,8.8~PeV$)=1.23$. Note that in the $\nu_e$ CC channel all the neutrino energy is transferred to the cascade, while for the NC ones the average inelasticity of the interaction is $\langle y \rangle\simeq 0.26$, and hence the cascades produced will have a lower energy, but for the range of neutrino energies considered they will still be above PeV energies. 
Note that the ratio of resonant to non-resonant rates further decreases for wider energy intervals or if one were to include the muon and tau channels.

We see that although the total peak cross section at $E_{GR}$ is 
$\sim 350$ times the value of $\sigma^{CC}$, taking all the relevant factors into account the final enhancement which can be obtained with this process for a wide neutrino spectrum around $E_{GR}$  is modest (of order unity)\footnote{While this work was being finished, a preprint appeared discussing possible signatures of the Glashow resonance and non-standard scenarios to account for the two IceCube events \cite{bh12}.}.

\section{Conclusions}

We have discussed the impact of the different production mechanisms of cosmogenic neutrinos on the fluxes at PeV energies. The main results are that:

- The neutrinos resulting from decays of neutrons produced in photo-pion interactions off the CMB are at a level $E_\nu^2 $d$\Phi_\nu/$d$E  < 10^{-10} $GeV/cm$^{2}$~s~sr
at PeV energies, and hence make a negligible contribution.

- The neutrinos resulting from decays of neutrons produced by photodisintegration of heavy nuclei are at a level $E_\nu^2 $d$\Phi_\nu/$d$E <  10^{-11} $GeV/cm$^{2}$~s~sr, and are hence also negligible.

- The neutrinos  resulting from pion decays produced in photo-pion interactions with UV/optical/IR photons are the dominant ones at PeV energies, and can reach the level of $E_\nu^2 $d$\Phi_\nu/$d$E \sim  10^{-9} $GeV/cm$^{2}$~s~sr in proton scenarios, and somewhat lower in heavy nuclei scenarios.

- The PeV neutrino fluxes  may be enhanced in scenarios with strong source redshift evolution or steeper source spectra, and also if the UV background were strongly enhanced at high redshift. 

- The overall contribution to the observable rates in the 1--10 PeV range due to the Glashow resonance is not large for the cosmogenic neutrinos. For instance, the $\bar\nu_e e$ resonant interaction leads to a contribution to the  rates of cascade events comparable to the non-resonant one resulting  from the neutrinos within 2.5~PeV of $E_{GR}$.

When taking into account the CR primary spectrum as well as the secondary gamma ray cascade  flux it turns out to be difficult to interpret the two IceCube events in terms of a cosmogenic neutrino flux, unless the two events were an upward fluctuation of the expected neutrino rates. Hence, the interpretation in terms of neutrino production at the sources seems likely, and these need not be necessarily at high redshifts.
 
\section*{Acknowledgments}
 We thank H. Asorey for assistance,  P. Schiffer and J. Kulbartz for discussing the role of the IRB and B. Sarkar for suggestions of simulation parameters.
The work of S. M. and E. R. is supported by CONICET PIP 112-200801-01830 and ANPCyT PICT 2010 1531, Argentina. The work of G. S. and A. vV. is supported by the Deutsche Forschungsgemeinschaft through
the collaborative research centre SFB 676, by BMBF under grant
05A11GU1, by the ``Helmholtz Alliance for Astroparticle Physics
(HAP)'' funded by the Initiative and Networking Fund of the Helmholtz
Association and by the State of
Hamburg, through the Collaborative Research program ``Connecting
Particles with the Cosmos''. We thank the Pierre Auger Collaboration 
for permission to use their data prior to journal publication.

\end{document}